\documentclass[draft]{cmmp}
\renewcommand{\author}[1]{\vskip-9\bigskipamount\centerline{\em #1}\vfill}
\begin{document}
\chapter{Black Holes in the Real Universe and Their Prospects as Probes of 
Relativistic Gravity}
\vspace{1.5 truecm}

\author{Martin J. Rees}

\section{Introduction}
Collapsed objects have definitely been observed:  
some are stellar-mass objects, 
the endpoint of massive stars; others, millions of times more massive, 
have been discovered  in the cores of most galaxies. Their formation  poses 
some  still-unanswered questions. But for relativists the key question is 
whether  observations can  probe the metric in the
strong-field domain,  and test whether it indeed agrees with the Kerr
geometry predicted by general relativity.

 In a lecture in 1977, Chandrasekhar wrote:

\begin{quote}``In my entire scientific life the most shattering experience has been
the realisation that exact solutions to Einstein's theory [the Kerr
solutions] provide the {\it absolutely exact representation} of untold
numbers of massive black holes that populate the Universe.''\end{quote}

He was of course referring to the uniqueness theorems, to which Stephen Hawking and several other participants at this conference contributed. At the time that Chandra
wrote this, there was some evidence for black holes. Now the evidence
is much stronger and my theme in this paper is to summarize this
evidence.

I will discuss first the black holes of a few solar masses that are
the remnants of supernovae, and then supermassive black holes in the
mass range of a million to a billion solar masses, for which there is
also good evidence. I shall also mention parenthetically that there
may be a population of black holes in the range of hundreds of solar
masses, relics of an earlier generation of massive stars. Bernard
Carr's paper discusses primordial black holes, right down to very low
masses, for which I think there is no evidence but which would be
fascinating if they existed.

      By the early  1970s   the theoretical  properties   of 
black holes -- the Kerr metric, the `no hair' theorems, and so forth -- 
were well established. But it took much longer for the observers to 
discover candidate black holes, and to recognise their nature. A 
fascinating review of the history has been given by Werner Israel [1]. By 
now, the evidence points insistently towards the presence of dark objects, 
with deep potential wells  and `horizons' through which matter can  pass 
into invisibility; however, the evidence does not yet allow observers to 
confirm the form of the metric in the strong-gravity region. I shall 
briefly summarise the evidence, and then comment on the prospects for 
testing strong-field gravity by future observations.

\section{Stellar mass holes}
      The first black hole candidates  to be identified  were bodies of a 
few solar masses, in close orbits around  an ordinary companion star, 
emitting intense and rapidly flickering  X-rays. This emission is 
attributed to inward-spiralling material, captured from the companion star, 
which swirls inward towards a `horizon'.  There are two categories of such 
binaries: those where the companion star is of high mass, of which Cygnus X1 
is the prototype, and the low-mass X-ray binaries (LMXBs) where the 
companion is typically below a solar mass. The LMXBs are sometimes called 
`X-ray novae' because they flare up to high luminosity: they plainly have a 
different evolutionary history from systems like Cygnus X1.
    
The X-ray emission from these objects was distinctively different from 
the periodic variability of a related class, which were believed to be 
neutron stars. The masses can be inferred from straightforward Newtonian 
arguments.  It is gratifying that the masses inferred for the periodic 
sources are in all cases below 2 solar masses (consistent with the 
theoretically-expected mass range for neutron stars) whereas those that 
vary irregularly have higher masses. There are now a dozen or so strong 
candidates of this type. Of course, the only stellar-mass holes that 
manifest themselves conspicuously are the tiny and atypical fraction that 
lie in close binaries where mass transfer is currently going on. There may 
be only a few dozen such systems in our Galaxy. But there are likely to be 
at least $10^7$ black holes in our Galaxy. This number is based on the rather 
conservative estimate that only 1 or 2 percent of supernovae leave black 
holes rather than neutron stars.  There could be a further population of 
black holes (maybe in the Galactic Halo) as a relic of early galactic 
history.[2]

Gamma ray bursts are a class of objects, of which about 3000 have been
detected, which flicker on time scales of less than a second and last
sometimes for only a second; sometimes for a few minutes. They are
known to be at great distances:  indeed they are so bright that if a
gamma-ray burst went off in our galaxy it would be as bright as the
Sun for a few seconds. They can be detected by X-ray or gamma-ray
telescopes even at very large redshifts. These are probably a rare
kind of collapsing star or merger of a compact binary which may signal
the formation of a black hole. They are extremely important because
they may be objects where we see a non-stationary black hole.

\section{Supermassive holes}
      A seemingly quite distinct class of black hole candidates  lurk in 
the centres of most galaxies; they are  implicated in the power output from
active galactic nuclei (AGNs), and in the production of relativistic jets
that energise strong radio sources. The demography of these massive holes
has been clarified by studies of relatively nearby galaxies: the centres of
most of these  galaxies display  either no activity or a rather low level,
but most  seem to harbour  dark central masses.

In most cases the evidence is of two kinds.  Stars in the innermost parts 
of  some galaxies are moving anomalously fast, as though `feeling;' the 
gravitational pull of a dark central mass; and in some galaxies with 
active nuclei, the central mass can be inferred by modelling the properties 
of the gas which reprocesses central continuum radiation  into  emission 
with spectral features. But there are two galaxies where central dark 
masses are indicated by other kinds of evidence that are far firmer.

      The first, in the peculiar spiral  NGC 4258,
 was revealed by  amazingly precise mapping of gas motions via the 1.3
cm maser emission line of ${\rm H}_2$O [3].  The spectral resolution of 
this
microwave line is high enough to pin down the velocities with accuracy of 1
km/sec.  The Very Long Baseline Array achieves an angular resolution better
than  0.5 milliarc seconds (100 times sharper than the HST, as well as far
finer spectral resolution of velocities!). These observations have
revealed, right in NGC 4258's core, a  disc with rotational speeds
following an exact Keplerian law around a compact dark mass.   The inner
edge of the observed disc  is orbiting at 1080 km/sec. It would be
impossible to circumscribe, within its radius, a stable and long-lived star
cluster with the inferred mass of $3.6 .10^7 \, {\hbox{$\rm\thinspace 
M_{\odot}$}}$.

The second utterly convincing candidate lies in our own Galactic Centre 
(see [4] for a comprehensive review). Direct 
evidence used to be ambiguous
because intervening gas and dust in the plane of the Milky Way prevents us
from getting a clear optical view of the central stars, as we can in, for
instance, M31. A great deal was known about gas motions, from radio and
infrared measurements, but these were hard to interpret because gas  does
not move ballistically like stars, but can be influenced by pressure
gradients, stellar winds, and other non-gravitational influences.
There is now, however,  direct evidence from
stellar proper motions, observed in the near infrared band, where
obscuration by intervening material is less of an obstacle [5,6].  The 
speeds
scale as $r^{-1/2}$ with distance from the centre, consistent with a hole of
mass $2.6 .10^6 \,{\hbox{$\rm\thinspace 
M_{\odot}$}}$. One can actually plot out the orbits of these stars. Some will make a complete circuit in little more than 100 years.  Corroboration  comes from the  compact  radio 
source that has
long been known to exist right at the dynamical centre of our Galaxy, which
can be interpreted in terms of accretion onto a massive hole  [7,8]. 

There is a remarkably close  proportionality    [9]  between the hole's
mass  and the velocity dispersion  of the central bulge or spheroid in the
stellar distribution (which is of course the dominant  part of an
elliptical galaxy, but only a subsidiary component of a disc system like
M31 or our own Galaxy).

\section{Scenarios for Black Hole Formation}

    There is no mystery about why high-mass stars may yield gravitationally 
collapsed remnants of 10 solar masses or more, but the formation route for 
the supermassive holes  is still uncertain.
Back in 1978, I presented a  `flow diagram' [10] exhibiting several  evolutionary 
tracks within a galaxy, aiming to convey the message that it seemed  almost 
inevitable that large masses
would collapse in galactic centres. 
There was not yet (see [1]) any consensus  that active 
galactic nuclei were powered by black holes (despite earlier arguments by 
Salpeter [11], Zeldovich and Novikov [12] and, especially,  Lynden Bell 
[13])   We have now got used to the idea that 
black holes indeed exist within most galaxies, but it is rather depressing 
that we still cannot decide which formation route is most likely.

     The  main options are summarised below:

\subsection{Monolithic formation of supermassive objects}
One possibility is that the gas in a newly-forming galaxy  does not all
break up into stellar-mass condensations, but that some  undergoes
monolithic collapse. As the gas evolves  (through loss of energy and angular
momentum) to a state of higher densities and more violent internal
dissipation, radiation pressure would prevent fragmentation, and   puff it
up into a single superstar.  Once a large mass of gas started to behave 
like a single superstar, it would continue to contract and deflate. Some 
mass would inevitably be
shed, carrying away angular momentum, but the remainder  would undergo
complete gravitational collapse. The behaviour of supermassive stars was 
studied by Bardeen, Thorne and others in the 1960s. Because radiation 
pressure is overwhelmingly dominant, such objects are destabilised, in the 
post-Newtonian approximation, even when  hundreds of times larger than the 
gravitational radius.  Rotation has a stabilising effect, but there has 
still been little work done on realistic models with differential rotation 
(see, however [14])

     The mass of the hole would depend on that of its host galaxy,
though not necessarily via an exact proportionality: the  angular momentum
of the protogalaxy and the depth of its central  potential well are
relevant factors too.
\subsection{Mergers of smaller holes (stellar mass or `intermediate mass')}
Rather than forming `in one go' from a superstar that may already be a 
million solar masses or more, holes might grow from smaller  beginnings. 
The first-generation stars are thought to have been more massive than those 
forming in galaxies today -- perhaps up to hundreds of solar masses. Such 
stars live no more than a few million years, and leave black hole
remnants if they lie in two distinct mass ranges [15]:

(i) Ordinary massive stars, with helium core masses up to $64 \, {\hbox{$\rm\thinspace 
M_{\odot}$}}$; and 
\hfill\break
(ii) `Very
Massive Objects' (VMOs) with helium cores above $130 \,{\hbox{$\rm\thinspace 
M_{\odot}$}}$.
\hfill\break
Stars in between
these two ranges leave no compact remnant at all, instead ending  their
lives by a disruptive explosion induced by the onset of electron-positron
pair production.

VMO
remnants  could have interesting implications in the present
universe [2];  such objects,
captured by supermassive holes, would yield gravitational radiation signals
detectable by LISA out to redshifts of order unity, possibly dominating the
event rate.
 Could there be a link between these `intermediate mass' holes and
supermassive holes? There are two possibilities. The most obvious, at first
sight, is that a cluster of such objects might merge into one. But it is
not easy  for a cluster of black holes to merge into a single one.  To see
this,   note that  one binary  with orbital speed $10^4$ km/sec (with a 
separation  of  1000 Schwarzschild radii)  would have just as
much binding energy as a cluster of 10000 holes with velocity dispersion
100 km/sec. Thus, if a cluster accumulated in the centre of a galaxy, the
likely outcome would be the expulsion of most objects, as the consequence
of  straightforward N-body dynamics, leaving only a few.

The prospects  of build-up by this route are not quite as bad as this
simple argument suggests, because the binding energy of the compact cluster
could be enhanced by dynamical friction on lower-mass stars, by gas drag,
or by gravitational radiation. This nonetheless seems an inefficient route
towards supermassive holes.

Ordinary stars, with large geometrical cross-sections, have a
larger chance of sticking together than pairs of black holes. We therefore 
cannot exclude  a  `scenario' where a supermassive star builds up  within a 
dense central cluster of ordinary stars.  The most detailed calculations 
were done by Quinlan and Shapiro ([16] and other references cited therein). 
These authors showed that stellar
coalescence,  followed by the segregation of the resultant high-mass stars
towards the centre, could  trigger runaway evolution without (as earlier
and cruder work had suggested) requiring clusters whose initial parameters
were unrealistic (i.e. already extremely dense, or with implausibly high
velocity dispersions).  It would be well worthwhile extending these
simulations to a wider range of initial conditions, and also to follow the
build-up from stellar masses to supermassive objects.

\subsection{Runaway growth of a favoured  stellar-mass hole  to 
supermassive status}
Even if a large population of low-mass holes is unlikely to merge
together, is it, alternatively, possible for one  of them, in a specially
favoured high-density environment, to undergo runaway growth via accretion?
An often-cited  constraint on the growth rate is based on the  argument
that,  however high  the external density was,  growth could not happen on
a timescale less than  the classic `Salpeter time'[11]

\begin{equation}
t_{\rm Sal} = 4 \times 10^7 (\varepsilon / 0.1) \, {\rm yrs}
\end{equation}

   For an efficiency $\varepsilon$ of 0.1 this would yield an e-folding 
timescale of
$4 \times 10^7$ years. If these holes started off with stellar masses, or 
as the
remnants of Population III stars,  there would seem to be barely enough
time for them to  have grown fast enough to energise quasars at $z = 6$,

This is not, however, a generic constraint;  there
are  several  suggestions in the literature for evading it. [17-20]

\section{The Galactic Context}
\subsection{The key issues}
Physical conditions in the central potential wells of young and
gas-rich galaxies should be propitious for black hole formation:  such
processes, occurring in the early-forming galaxies that develop from
high-amplitude peaks in the initial density distribution,  are presumably
connected with high-$z$ quasars.   It now seems clear that most galaxies 
that
existed at $z= 3$ would have participated subsequently in a series of
mergers;  giant  present-day elliptical galaxies are the outcome of such
mergers.   Any black holes already present would tend to spiral inwards,
and  coalesce [21,22] (unless a third body fell in before the merger was 
complete,
in which case  a Newtonian  slingshot could eject all three:  a binary in
one direction; the third, via recoil, in the opposite direction).

The issues  for astrophysicists are  then:\hfill\break
 (a)  How much does a black hole grow by gaseous accretion  (and how much
electromagnetic energy  does it radiate) at each stage? Models based on
semi-analytic schemes for galaxy evolution have achieved a good fit with
the luminosity function and $z$-dependence of quasars.  Less gas is 
available
at later epochs, and this accounts for the scarcity of high-luminosity AGNs
at low $z$.\hfill\break
and\hfill\break
 (b) How far back along the `merger tree'  did this process start? A
single big galaxy can be traced back to the stage when it was in hundreds
of smaller components with individual internal velocity dispersions  as low
as  20 km/sec. Did central black holes form  even in these small and weakly
bound systems?
 This issue has been widely discussed (see for instance [23-26] and 
references cited therein). It is important because it determines
whether there is a population of high-$z$ miniquasars.

\subsection{Tidally-disrupted stars}
When the central hole mass is below $10^8 \, {\hbox{$\rm\thinspace 
M_{\odot}$}}$, solar-type stars are
disrupted before they get close enough to fall within the hole's horizon. A
tidally disrupted star, as it moves away from the hole, develops into an
elongated banana-shaped structure, the most tightly bound debris (the first
to return to the hole) being at one end.  There would not be a conspicuous
`prompt' flare signalling the disruption event, because the energy
liberated is trapped within the debris.  Much more radiation emerges when
the bound debris (by then more diffuse and transparent)  falls back onto
the hole a few months later,  after completing an eccentric orbit. The
dynamics and radiative transfer are then even more complex and uncertain
than in the disruption event itself, being affected by relativistic
precession, as well as by the effects of viscosity and shocks.

The radiation from the inward-swirling debris would be predominantly
thermal, with a temperature of order $10^5$   K;  however the energy 
dissipated
by the shocks that occur during the circularisation  would provide an
extension into the X-ray band.  High luminosities would be attained -- the
total photon energy radiated (up to $10^{53}$ ergs) could be several 
thousand times more than the  {\it photon}   output of a supernova. The 
flares would,
however, not be standardised -- what is observed would depend on the hole's
mass and spin, the type of star, the impact parameter, and the orbital
orientation relative to the hole's spin axis and the line of sight; perhaps
also on absorption in the galaxy. To compute what happens involves
relativistic gas dynamics and radiative transfer, in an unsteady  flow with
large dynamic range, which possesses no special symmetry and therefore
requires full 3-D calculations -- still a  daunting  computational
challenge. [27-30]

    The rate of tidal disruptions in our Galactic Centre would be no more
than once per $10^5$ years.   But each such event could generate a 
luminosity
several times $10^{44}$ erg/s   for about a year. Were this in the UV, the 
photon
output, spread over $10^5$  years, could exceed the current ionization rate:
the mean output might exceed the median output.  The radiation emitted from 
the event might reach us after a delay
if it  reflected or flouresced  off surrounding material. Sunyaev and his
collaborators have already  used such considerations  to set  non-trivial
constraints on the history of the Galactic Centre's X-ray output over the
last few thousand years.
\section{Do the Candidate Holes  Obey the Kerr Metric?}

\subsection{Probing  near the hole}

     The observed molecular disc in NGC 4258 lies a long way out: at
around $10^5$ gravitational radii.  We can exclude all conventional
alternatives (dense star clusters, etc); however, the measurements
tell us nothing about the central region where gravity is strong --
certainly not whether the putative hole actually has properties
consistent with the Kerr metric. The stars closest to our Galactic
Centre likewise lie so far out from the putative hole (their speeds
are less than 1 percent that of light) that their orbits are
essentially Newtonian.

       We can infer from AGNs that `gravitational pits' exist, which
must be deep enough to allow several percent of the rest mass of
infalling material to be radiated from a region compact enough to vary
on timescales as short as an hour. But we still lack quantitative
probes of the relativistic region. We believe in general relativity
primarily because it has been resoundingly vindicated in the weak
field limit (by high-precision observations in the Solar System, and
of the binary pulsar) -- not because we have evidence for black holes
with the precise Kerr metric.

    The emission from most accretion flows is concentrated towards the 
centre, where the potential well is
deepest and the motions fastest.  Such basic features of the
phenomenon as the overall efficiency, the minimum variability
timescale, and the possible extraction of energy from the hole itself
all depend on inherently relativistic features of the metric -- on
whether the hole is spinning or not, how it is aligned, etc.  But the
data here are imprecise and `messy'.  We would occasionally expect to
observe, even in quiescent nuclei, the tidal disruption of a
star. Exactly how this happens would depend on distinctive precession
effects around a Kerr metric, but the gas dynamics are so complex that
even when a flare is detected it will not serve as a useful diagnostic
of the metric in the strong-field domain. There are however several
encouraging new possibilities.

 \subsection{X-ray  spectroscopy of accretion flows}

       Optical spectroscopy tells us a great deal about the gas in
AGNs.  However, the optical spectrum originates quite far from the
hole. This is because the innermost regions would be so hot that their
thermal emission emerges as more energetic quanta.  X-rays are a far
more direct probe of the relativistic region.  The appearance of the
inner disc around a hole, taking doppler and gravitational shifts into
account, along with light bending, was first calculated by Bardeen and
Cunningham [31] and subsequently by several others (eg [32]). There is
of course no short-term hope  of
actually `imaging' these inner discs (Though an X-ray interferometer called MAXIM, with elements separated by 500 km,  is being studied.) 
However,  we need not wait that long for a probe, because the large
frequency-shifts 
predicted in (for instance) the 6.4 keV line from Fe
could reveal themselves spectroscopically --
substantial gravitational redshifts would be expected, as well as
large doppler shifts [33].  Until recently, the energy resolution and
sensitivity of X-ray detectors was inadequate to permit spectroscopy
of extragalactic objects. The ASCA X-ray satellite was the first with
the capability to measure emission line profiles in AGNs.  There is
already one convincing case [34] of a broad asymmetric emission line
indicative of a relativistic disc, and others should soon follow.  The
value of (a/m) can in principle be constrained too, because the
emission is concentrated closer in, and so displays larger shifts, if
the hole is rapidly rotating, and there is some evidence that this
must be the case in two objects [35, 36].

   The recently-launched Chandra and XMM X-ray satellites are now able to extend and
refine these studies; they may offer enough sensitivity, in combination
with time-resolution, to study flares, and even to follow a `hot spot'
on a plunging orbit.

   The swing in the polarization vector of photon trajectories near a
hole was long ago suggested [37] as another diagnostic; but this is
still not feasible because X-ray polarimeters are far from capable of
detecting the few percent polarization expected.

\subsection{The Blandford-Znajek process}

    Back in 1969 Penrose [38] showed how energy could in principle be 
extracted from a spinning hole. Some years later Blandford and Znajek [39] 
proposed an astrophysically-realistic  process whereby this might happen: 
a magnetic field threading a
hole (maintained by external currents in, for instance, a torus) could
extract spin energy, converting it into directed Poynting flux and
electron-positron pairs.

     Can we point to objects where this is
definitively happening?  The giant radio lobes from radio galaxies
sometimes spread across millions of lightyears -- $10^{10}$ times larger
than the hole itself. If the Blandford-Znajek process is really going
on, these huge structures may be the most direct manifestation of an
inherently relativistic effect around a Kerr hole.

    Jets in some AGNs definitely have Lorentz factors exceeding
10. Moreover, some are probably Poynting-dominated, and contain electron-positron
(rather than electron-ion) plasma. But there is still no compelling
reason to believe that these jets are energised by the hole itself,
rather than by winds and magnetic flux `spun off' the surrounding
torus. The case for the Blandford-Znajek mechanism would be
strengthened if baryon-free jets were found with still higher Lorentz
factors, or if the spin of the holes could be independently measured,
and the properties of jets turned out to depend on (a/m).

  The process cannot dominate unless either the field threading the
hole is comparable with that in the orbiting material, or else the
surrounding material radiates with low radiative efficiency. These
requirements cannot be ruled out, though there has been recent
controversy about how plausible they are. (The Blandford-Znajek effect could  
be important in the still more
extreme context of gamma-ray bursts, where a newly formed hole of a
few solar masses could be threaded by a field exceeding $10^{15}$  G.)

\subsection{What is the expected spin?}

  The spin of a hole affects the efficiency of `classical' accretion
processes; the value of a/m also determines how much energy is in
principle extractable by the Blandford-Znajek effect.  Moreover, the
orientation of the spin axis may be important in relation to jet
production, etc.

    Spin-up is a natural consequence of prolonged disc-mode accretion:
any hole that has (for instance) doubled its mass by capturing
material that is all spinning the same way would end up with a/m being
at least 0.5.  A hole that is the outcome of a merger between two of
comparable mass would also, generically, have a substantial spin. On
the other hand, if it had gained its mass from capturing many low-mass
objects (holes, or even stars) in randomly-oriented orbits, a/m would
be small.

\subsection{Precession and alignment}

   Most of the  extensive literature on
gas dynamics around Kerr holes assumes that the
flow is axisymmetric. This assumption is motivated not just by
simplicity, but by the expectation that Lense-Thirring precession
would impose axisymmetry close in, even if the flow further out were
oblique and/or on eccentric orbits.  Plausible-seeming arguments,
dating back to the pioneering 1975 paper by Bardeen and Petterson [40],
suggested that the alignment would occur, and would extend out to a
larger radius if the viscosity were low because there would be more
time for Lense-Thirring precession to act on inward-spiralling gas.
However, later studies, especially by Pringle, Ogilvie, and their
associates, have shown that naive intuitions can go badly awry. The
behaviour of the `tilt' is much more subtle; the effective viscosity
perpendicular to the disc plane can be much larger than in the
plane. In a thin disc, the alignment effect is actually weaker when
viscosity is low. What happens in a thick torus is a still unclear,
and will have to await 3-D gas-dynamical simulations.

        The orientation of a hole's spin and the innermost flow
patterns could have implications for jet alignment. An important paper
by Pringle and Natarajan [41] shows that `forced precession' effects
due to torques on a disc can lead to swings in the rotation axis that
are surprisingly fast (i.e. on timescales very much shorter than the
timescale for changes in the hole's mass).

\subsection{Stars in relativistic orbits?}

    Gas-dynamical phenomena are complicated because of viscosity,
magnetic fields etc. It would be nice to have a `cleaner' and more
quantitative probe of the strong-field regime: for instance, a small
star orbiting close to a supermassive hole.  Such a star would behave
like a test particle, and its precession would probe the metric in the
`strong field' domain.  These interesting relativistic effects, have
been computed in detail by Karas and Vokrouhlicky [42.43]. Would we
expect to find a star in such an orbit?

   An ordinary star certainly cannot get there by the kind of `tidal
capture' process that can create close binary star systems. This is
because the binding energy of the final orbit (a circular orbit with
the same angular momentum as an initially near-parabolic orbit with
pericentre at the tidal-disruption radius) would have to be dissipated
within the star, and that cannot happen without destroying it.  An orbit 
can however  be
`ground down' by successive impacts on a disc (or any other resisting
medium) without being destroyed [44]: the orbital energy then goes almost
entirely into the material knocked out of the disc, rather than into
the star itself. And there are other constraints on the survival of stars 
in the
hostile environment around massive black holes -- tidal dissipation
when the orbit is eccentric, irradiation by ambient radiation, etc. [45,46].
They can be thought of as close binary star systems with extreme mass
ratios.

   These stars would not be directly observable, except maybe in our
own Galactic Centre.  But they might have indirect effects: such a
rapidly-orbiting star in an active galactic nucleus could signal its
presence by quasiperiodically modulating the AGN emission.

\subsection{Gravitational-wave capture of compact stars}

   Neutron stars or white dwarfs circling close to supermassive black
holes would be impervious to tidal dissipation, and would have such a
small geometrical cross section that the `grinding down' process would
be ineffective too. On the other hand, because they are small they can
get into very tight orbits by straightforward stellar-dynamical
processes. For ordinary stars, the `point mass' approximation breaks
down for encounter speeds above 1000 km/s -- physical collisions are
then more probable than large-angle deflections. But there is no
reason why a `cusp' of tightly bound {\it compact} stars should not extend
much closer to the hole.  Neutron stars or white dwarfs could exchange
orbital energy by close encounters with each other until some got
close enough that they either fell directly into the hole, or until
gravitational radiation became the dominant energy loss. When stars
get very close in, gravitational radiation losses become significant, and
tend to circularise an elliptical orbit with small pericentre. Most
such stars would be swallowed by the hole before circularisation, because 
the
angular momentum of a highly eccentric orbit `diffuses' faster than
the energy does due to encounters with other stars, but some would get
into close circular orbits [47,48].

    A compact star is less likely than an ordinary star in similar
orbit to `modulate' the observed radiation in a detectable way.  But
the gravitational radiation (almost periodic because the dissipation
timescale involves a factor $(M_{\rm hole}/m^*)$) would be detectable.

\subsection{Scaling laws and `microquasars'}

     Two galactic X-ray sources that are believed to involve black
holes generate double radio structures that resemble miniature
versions of the classical extragalactic strong radio sources. The jets have 
been found to
display apparent superluminal motions across the sky, indicating that,
like the extragalactic radio sources, they contain plasma that is
moving relativistically. [49]

  There is no reason to be surprised by this analogy between phenomena
on very different scales. Indeed, the physics is exactly the same,
apart from very simple scaling laws.
If we define $l = L/L_{\rm Ed}$ and $\dot m = \dot M/
\dot M_{\rm crit}$, where $\dot M_{\rm crit} = L_{\rm Ed}/c^2$, then for a 
given value of $\dot m$, the flow
pattern may be essentially independent of $M$. Linear scales and
timescales, of course, are proportional to M, and densities in the
flow scale as $M^{-1}$.  The physics that amplifies and tangles any
magnetic field may be scale-independent, and the field strength B
scales as $M^{-1/2}$. So the bremsstrahlung or synchrotron cooling
timescales go as $M$, implying that $t_{\rm cool}/t_{\rm dyn}$ is 
insensitive to $M$ for a given $\dot m$. So
also are ratios involving, for instance, coupling of electron and ions
in thermal plasma. Therefore, the efficiencies and the value of $l$ are
insensitive to $M$, and depend only on $\dot m$.  Moreover, the form of the
spectrum, for given $\dot m$,  depends on M only rather insensitively (and 
in a manner that
is easily calculated).

  The kinds of accretion flow inferred in, for instance, the centre of the giant galaxy M87, giving
rise to a compact radio and X-ray source, along with a relativistic
jet, could operate just as well if the hole mass was lower by a
hundred million, as in the galactic  sources.  So we can actually
study the same processes involved in AGNs in microquasars close at
hand within our own galaxy. And these miniature sources may allow us
to observe, in our own lifetimes, a simulacrum of the entire evolution of a strong
extragalactic radio source, its life-cycle speeded up by a similar factor.

\subsection{Discoseismology}

   Discs or tori that are maintained by steady flow into a black hole
can support vibrational modes [50-52].  The frequencies of these modes
can, as in stars, serve as a probe for the structure of the inner disc
or torus. The amplitude depends on the importance of pressure, and
hence on disc thickness; how they are excited, and the amplitude they
may reach, depends, as in the Sun, on interaction with convective
cells and other macroscopic motions superimposed on the mean flow. But
the frequencies of the modes can be calculated more reliably. In
particular, the lowest g-mode frequency is close to the maximum value
of the radial epicyclic frequency $k$.  This epicyclic frequency is, in
the Newtonian domain, equal to the orbital frequency. It drops to zero
at the innermost stable orbit. It has a maximum at about $9GM/c^2$ for a
Schwarzschild hole; for a Kerr hole, $k$ peaks at a smaller radius (and
a higher frequency for a given $M$). The frequency is 3.5 times higher
for $(a/m)=1$ than for the Schwarzschild case.

   Novak and Wagoner [52] pointed out that these modes may cause an
observable modulation in the X-ray emission from galactic black hole
candidates.  Just such quasi-periodicities  have been seen. The amplitude 
is a few percent (somewhat larger at harder X-ray
energies) suggesting that the oscillations involve primarily the
hotter inner part of the disc. In one object, known as  GRS 1915+105
(a galactic object which also emits relativistic jets), the 
fluctuation spectrum showed a peak
in Fourier space at around 67 Hz. This frequency does not change even
when the X-ray luminosity doubles, suggesting that it relates to a
particular radius in the disc. If this is indeed the lowest g-mode,
and if the simple disc models are relevant, then the implied mass is
$10.2 \, {\hbox{$\rm\thinspace 
M_{\odot}$}}$ for Schwarzschild, and $35 \, {\hbox{$\rm\thinspace 
M_{\odot}$}}$ for a `maximal Kerr' hole 
[52]. 
Several other X-ray sources have been found to display quasi-periodicities, and other regularities: for instance, two superimposed frequencies which change, but in such a way that the difference between them is constant. If such regularities can be understood, they offer 
 the exciting prospect of inferring (a/m) for holes
whose masses are independently known.

\section{Gravitational Radiation as a Probe}

\subsection{Gravitational waves from newly-forming massive holes?}

     The gravitational radiation from black holes offers impressive
tests of general relativity, involving no physics other than that of
spacetime itself.

     At first sight, the formation of a massive hole from a monolithic
collapse might seem an obvious source of strong wave pulses.  The
wave emission would be maximally intense and efficient if the holes
formed on a timescale as short as $(r_g/c)$, where $r_g= (GM/c^2)$ --
something that might happen if they built up via coalescence of
smaller holes (cf ref [21]).

       If, on the other hand, supermassive black holes formed from 
collapse of an unstable supermassive star, then the process may be too 
gradual to yield efficient gravitational radiation.  That is because 
post-Newtonian
instability is triggered at a radius $r_i \gg r_g$.  Supermassive stars
are fragile because of the dominance of radiation pressure: this
renders the adiabatic index only slightly above 4/3 (by an amount of
order $10 ^{-1/2})\, {\hbox{$\rm\thinspace 
M_{\odot}$}}$. Since = 4/3 yields neutral stability in 
Newtonian
theory, even the small post-Newtonian corrections then destabilize
such `superstars'.  The characteristic collapse timescale when
instability ensues is longer than $r_g/c$ by the 3/2 power of the
collapse factor.

    The post-Newtonian instability is suppressed by rotation.  A
differentially rotating supermassive star could in principle support
itself against post-Newtonian instability until it became very tightly
bound.  It could then perhaps develop a  bar-mode instability and
collapse within a few dynamical times.  cf ref [ 14]. To achieve this 
tightly-bound
state without drastic mass loss, the object would need to have deflated 
over a long timescale,
losing energy at no more than the Eddington rate.

            The gravitational waves associated with supermassive holes
would be concentrated in a frequency range around a millihertz -- too
low to be accessible to ground-based detectors, which lose sensitivity
below 100 Hz, owing to seismic and other background
noise. Space-based detectors are needed. 
There are firm plans, discussed further in the papers by Shutz and Thorne, for the  Laser 
Interferometric Space Array (LISA)
-- three  spacecraft on solar orbit, configured as a triangle, with
sides  of 5 million km long whose length is monitored by laser
interferometry.

\subsection{Gravitational waves from coalescing supermassive holes.}

   The strongest signals are expected when already-formed holes coalesce, 
as the aftermath of mergers of their host galaxies.  Many galaxies have 
experienced a merger since the epoch $z > 2$
when, according to `quasar demography' arguments  they
acquired central holes. When two massive holes spiral together, energy is 
carried away by
dynamical friction (leading, when the binary is `hard', to expulsion of
stars from the galaxy) and also by drag on gas.  Eventually the members of 
the binary get close enough for gravitational radiation (with a timescale 
proportional to the inverse 4th power of separation)  to take over and 
drive them towards coalescence. In their final coalescence, up to $\sim 10$ 
per
cent of their rest mass as a burst of gravitational radiation in a
timescale of only a few times $r_g/c$. These pulses would be so strong
that LISA could detect them with high signal-to-noise even from large
redshifts.  Whether such events happen often enough to be interesting
can to some extent be inferred from observations (we see many galaxies
in the process of coalescing), and from simulations of the
hierarchical clustering process whereby galaxies and other cosmic
structures form.  The merger rate of the large galaxies believed to harbour 
supermassive
holes: it is only about one event per century, even out to redshifts $z
= 4$.  However, big galaxies are probably the outcome of many
successive mergers.  We still have no direct
evidence -- nor firm theoretical clues -- on whether these small
galaxies harbour black holes (nor, if they do, of what the hole masses
typically are). However it is certainly possible that enough holes of
(say) $10^5 \, {\hbox{$\rm\thinspace 
M_{\odot}$}}$ lurk in small early-forming galaxies to yield, via
subsequent mergers, more than one event per year detectable by LISA [53].

      LISA is potentially so sensitive that it could detect the
nearly-periodic waves from stellar-mass objects orbiting a
$10^5 -
10^6 \, {\hbox{$\rm\thinspace 
M_{\odot}$}}$  hole, even at a range of a hundred Mpc, despite the $m^*/M_{\rm 
hole}$ factor
whereby the amplitude is reduced compared with the coalescence of two
objects of comparable mass $M_{\rm hole}$. The stars in the observed 
`cusps' around
massive central holes in nearby galaxies are of course (unless almost
exactly radial) on orbits that are far too large to display
relativistic effects.  Occasional captures into relativistic orbits
can come about by dissipative processes -- for instance, interaction
with a massive disc [44]. But unless the hole mass were above $10^8 \, {\hbox{$\rm\thinspace 
M_{\odot}$}}$
 (in which case the waves would be at too low a frequency for LISA
to detect), solar-type stars would be tidally disrupted before getting
into relativistic orbits. Interest therefore focuses on compact stars,
for which dissipation due to tidal effects or drag is less
effective. As already described [47,48], compact stars may get captured
as a result of gravitational radiation, which can gradually `grind
down' an eccentric orbit with close pericenter passage into a
nearly-circular relativistic orbit. The long quasi-periodic wave
trains from such objects, modulated by orbital precession (cf refs
[22,23]) in principle carries detailed information about the metric.

     The attraction of LISA as an `observatory' is that even
conservative assumptions lead to the prediction that a variety of
phenomena will be detected.  If there were many massive holes not
associated with galactic centres (not to mention other speculative
options such as cosmic strings), the event rate would be much
enhanced. Even without factoring on an `optimism factor' we can be
confident that LISA will harvest a rich stream of data.

\subsection{Gravitational-wave recoil}

     Is there any way of learning, before that date, something about
gravitational radiation?  The dynamics (and gravitational radiation)
when two holes merge has so far been computed only for cases of
special symmetry. The more general problem -- coalescence of two Kerr
holes with general orientations of their spin axes relative to the
orbital angular momentum -- is a `grand challenge' computational
project being tackled at the Einstein Institute in Potsdam, and at other 
centres. When this challenge has been met (and  one hopes it will not take 
all the time until LISA flies) we shall find out not only the
characteristic wave form of the radiation, but the recoil that arises
because there is a net emission of linear momentum.

   There would be a recoil due to the non-zero net linear momentum
carried away by gravitational waves in the coalescence. If the holes
have unequal masses, a preferred longitude in the orbital plane is
determined by the orbital phase at which the final plunge occurs.  For
spinning holes there may, additionally, be a rocket effect
perpendicular to the orbital plane, since the spins break the mirror
symmetry with respect to the orbital plane. [54]

   The recoil is a strong-field gravitational effect which depends
essentially on the lack of symmetry in the system.  It can therefore
only be properly calculated when fully 3-dimensional general
relativistic calculations are feasible.  The velocities arising from
these processes would be astrophysically interesting if they were
enough to dislodge the resultant hole from the centre of the merged
galaxy.
   The recoil might even be so violent that the merged hole breaks
loose from its galaxy and goes hurtling through intergalactic space.
This disconcerting thought should at least impress us with the reality
and `concreteness' of the extraordinary entities to whose  understanding
Stephen Hawking has contributed so much.

\section{Acknowledgements}\nonumber

I am grateful to several colleagues, especially  Mitch Begelman,
Roger Blandford, Andy Fabian,  and Martin Haehnelt for discussions and
collaboration on topics mentioned here.


\begin{thebibliography}{52.}
\addcontentsline{toc}{section}{References}

\bibitem{journ1} W. Israel: Foundations of Physics \textbf{26}, 595 (1996)

\bibitem{journ2} P. Madau, M.J. Rees: Astrophys. J. \textbf{551}, L27 (2001)

\bibitem{journ3} K. Miyoshi  et al: Nature \textbf{373}, 127 (1995)


\bibitem{journ4} F. Melia, H. Falcke: Ann. Rev. Astr. Astrophys. \textbf{39}, 309 
(2001). 

\bibitem{journ5} A. Eikart, R. Genzel, T. Ott, R. Schodel:
MNRAS  \textbf{331}, 917 (2002)

\bibitem{journ6} A.M. Gehz, M. Morris, E.E. Becklin, A. Tanner, T. Kremenck: Nature \textbf{407}, 349 (2002)

\bibitem{proc7} M.J. Rees: `The Compact Source at the Galactic Center'. In 
\emph{The Galactic Center}, ed. by  G. Riegler, R.D. Blandford  (AIP) pp. 
166--176 (1982)

\bibitem{journ8} R. Narayan, I. Yi, R. Mahadevan: Nature \textbf{374}, 623 
(1995)

\bibitem{proc9} D. Merritt, L. Ferrarese: Astrophys. J. \textbf{547}, 140 (2001).

\bibitem{journ10} M.J. Rees: Observatory \textbf{98}, 210, (1978)

\bibitem {journ11} E.E. Salpeter: Astrophys. J. \textbf{140}, 796 (1964)

\bibitem {journ12} Y.B. Zeldovich, I.D. Novikov: Sov Phys. Dok 158, 811 
(1964)
\bibitem {journ13} D. Lynden-Bell: Nature, \textbf{223}, 690 (1969)



\bibitem{journ14} T.W. Baumgarte,   S.L. Shapiro:  Astrophys. J. \textbf{526}, 941.
(1999)

\bibitem{journ15} A. Heger, S. Woosley: Astrophys. J. \textbf{567}, 232 (2002)

\bibitem{journ16} G.D. Quinlan, S.L. Shapiro: Astrophys. J. \textbf{356}, 
483 (1990)

\bibitem{journ17} M.C. Begelman: MNRAS \textbf{187}, 237 (1979)

\bibitem{journ18} M. Abramowicz, M. Jaroszynski, M. Sikora: Astron. 
Astrophys. \textbf{63}, 221 (1980)

\bibitem{journ19} M.C. Begelman: Astrophys. J. \textbf{568}, L97 (2002)

\bibitem{journ20} N. Shaviv: Astrophys. J. \textbf{494}, L193 (1998)

\bibitem{journ21}  T. Ebisuzaki, J. Makino, S.K. Okumura: Nature
\textbf{354}, 212 (1991)

\bibitem{journ22} M. Milosavljevic, D. Merritt: Astrophys. J. \textbf{563}, 34 (2001)

\bibitem{journ23} Z. Haiman, A. Loeb: Astrophys. J. \textbf{499}, 520 (1998)

\bibitem{journ24}  M. Haehnelt, G. Kaufmann: MNRAS \textbf{318}, 235 (2000)

 \bibitem{journ25}    K. Menon, Z. Haiman, V.K. Narayanan: Astrophys. J. \textbf{558}, 535 (2001)

\bibitem{journ26} Q. Yu, S. Tremaine: Astrophys. J. (in press) astro-ph 0203082

\bibitem{journ27} B. Carter, J.-P. Luminet: Astr. Astrophys. \textbf{121}, 97 (1983)

\bibitem{journ28} J.K. Canizzo, H.M. Lee, J. Goodman: Astrophys. J. 
\textbf{351}, 38 (1990)

\bibitem{journ29} D. Syer, A. Ulmer: MNRAS \textbf{306}, 35 (1999)

\bibitem{journ30 } J. Magorrian, S. Tremaine: MNRAS \textbf{309}, 447 (1999)

\bibitem{journ31} J. Bardeen, J. Cunningham:   Astrophys. J. \textbf{173}, L137 (1972)

\bibitem{journ32} K.P. Rauch, R.D. Blandford: Astrophys. J. \textbf{421}, 46 (1993)

\bibitem{journ33} N.E. White et al: MNRAS \textbf{238}, 729 (1989).




\bibitem{journ34} Y. Tanaka: et al:  Nature \textbf{375}, 659 (1995) 

\bibitem{journ35} K. Iwasawa et al:  MNRAS \textbf{306}, L191 (1999)

\bibitem{journ36} J Wilms et al: MNRAS \textbf{328}, L27 (2001)

\bibitem{journ37} P.A. Connors, T. Piran, R.F. Stark:   Astrophys. J. \textbf{235}, 
224 (1980)

\bibitem{journ38} R. Penrose: Rivista del Nuivo Cimento, Numero speciale, \textbf{1}, 252 
(1969)

\bibitem{journ39} R.D. Blandford, R.L. Znajek:  MNRAS \textbf{179}, 433 (1977)

\bibitem{journ40} J. Bardeen, J.A. Petterson:   Astrophys. J. \textbf{195}, L65 (1975)

\bibitem{journal41} P. Natarajan, J.E. Pringle: Astrophys. J.
\textbf{506}, 97 (1998)

\bibitem{journ42} V. Karas, D. Vokrouhlicky: MNRAS \textbf{265}, 365 (1993)

\bibitem{journ43} V. Karas, D. Vokrouhlicky: Astrophys. J. \textbf{422}, 208 (1994)

\bibitem{journ44} D. Syer, C.J.  Clarke, M.J. Rees:   MNRAS \textbf{250}, 505 (1991)

\bibitem{journ45}  P. Podsiadlowski, M.J. Rees: In `Evolution of X-ray 
binaries'
ed S Holt and C.Day (AIP) p403 (1994)

\bibitem{journ46} A.R. King, C. Done: MNRAS \textbf{264}, 388 (1993)

\bibitem{journ47} D. Hils, P.L.  Bender:   Astrophys. J. (Lett) \textbf{445}, L7
(1995)

\bibitem{journ48} S. Sigurdsson, M.J. Rees: MNRAS \textbf{284}, 318 (1997)

\bibitem{journ49} F. Mirabel, L.F. Rodriguez: Ann. Rev. Astr. Astrophys:
\textbf{37}, 409 (1999)

\bibitem{journ50} S. Kato, J. Fukui:   PASJ  \textbf{32}, 377 (1980)

\bibitem{journ51} M.A. Novak, R.V. Wagoner:  Astrophys. J. \textbf{393}, 697 (1992)

\bibitem{journ52} M.A. Novak, R.V. Wagoner:   Astrophys J. \textbf{418}, 187 (1993)

\bibitem{journ53} M.G. Haehnelt: MNRAS \textbf{269}, 199 (1994)

\bibitem{journ54}I.  Redmount, M.J. Rees: Comm. Astrophys. Sp. Phys \textbf{14}  
185 (1989)
\end{thebibliography}
\end{document}